\documentclass[prl,aps,floats,twocolumn]{revtex4}

\usepackage[dvips]{graphicx}
\usepackage{amssymb}
\usepackage{amsmath}
\usepackage{float}

\usepackage{color}

\begin{document}

\title{Symmetric Satellite Swarms and Choreographic Crystals}

\author{Latham Boyle$^{1}$, Jun Yong Khoo$^{1,2}$ and Kendrick Smith$^{1,3}$}

\affiliation{$^{1}$Perimeter Institute for Theoretical Physics, Waterloo,
  Ontario, Canada \\ $^{2}$ MIT Physics Department, Cambridge, MA, USA \\
  $^{3}$ Princeton University Astronomy Department, Princeton, NJ, USA}

\date{July 2014}
                            
\begin{abstract}
In this paper, we introduce a natural dynamical analogue of crystalline order, which we call choreographic order.  In an ordinary (static) crystal, a high degree of symmetry may be achieved through a careful arrangement of the fundamental repeated elements.  In the dynamical analogue, a high degree of symmetry may be achieved by having the fundamental elements perform a carefully choreographed dance.  For starters, we show how to construct and classify all symmetric satellite constellations.  Then we explain how to generalize these ideas to construct and classify choreographic crystals more broadly.  We introduce a quantity, called the ``choreography" of a given configuration.  We discuss the possibility that some (naturally occurring or artificial) many-body or condensed-matter systems may exhibit choreographic order, and suggest natural experimental signatures that could be used to identify and characterize such systems.  
\end{abstract}

\maketitle 

\section{Introduction: The 4 Satellite Orbit}  

Because they are such natural and beautiful structures, lattices and crystals appear throughout physics and mathematics, in many different (and often unexpected) ways \cite{AshcroftMermin, ConwaySloane, Kac, Borcherds}.  Here we introduce the idea of a choreographic crystal, a type of configuration that can be much more symmetrical than is revealed by a snapshot of it at any given time.  We study this idea from several different angles, and suggest experimental diffraction signatures to identify and characterize such choreographic systems in the lab, whether they are naturally occuring or artificially engineered.  

Let us start with a simple and beautiful example of a choreographic lattice.  We can find 
our way to this example by comparing the following two elementary problems.

The first problem is static: what is the most symmetrical arrangement of four points on the 2-sphere?  The solution is well known: the four points are the four vertices of a regular tetrahedron.  We can express the positions of these four points neatly in cartesian coordinates as follows: we start from the 8 vertices of the cube $\{\pm1,\pm1,\pm1\}$ and select the four vertices with an even number of minus signs; in other words, the positions $\vec{q}_{\alpha}$ of the four points ($\alpha=0,1,2,3$) have cartesian components $q_{\alpha}^{\,j}$ ($j=1,2,3$) given by
\begin{equation}
  \label{q_hat}
  q_{\alpha}^{\,j}=(-1)^{1+\delta_{0,\alpha}+\delta_{j,\alpha}}
\end{equation}
where $\delta_{a,b}$ is the Kronecker delta function.  

The second problem is a natural dynamical analogue of the first: let us now imagine that we let the points flow along the geodesics of the sphere ({\it i.e}\ the great circles), with angular velocities that are constant in time and all have equal magnitudes, like satellites in circular orbit around the Sun -- what is the most symmetrical configuration of four such satellite trajectories?  Once again, the answer may be neatly summarized in cartesian coordinates: we choose the four satellites to have trajectories $\vec{p}_{\alpha}(t)$ with cartesian components $p_{\alpha}^{j}(t)$ given by
\begin{equation} 
  \label{p_hat}
  p_{\alpha}^{j}(t)=q_{\alpha}^{j}\;{\rm cos}\left(t-\frac{2\pi j}{3}\right)
\end{equation}
where $q_{\alpha}^{\;\!j}$ is given by Eq.~(\ref{q_hat}).  This solution has the following geometrical interpretation.   Each of the four satellites is orbiting in a different orbital plane: the $\alpha$th trajectory $\vec{p}_{\alpha}(t)$ is a circular orbit with its angular momentum in the direction $\vec{q}_{\alpha}$; in other words, there is one satellite orbiting around each of the four 3-fold symmetry axes of the regular tetrahedron.  Furthermore, to achieve maximal symmetry, the relative phases of the four orbits (or, equivalently, the initial positions) have been carefully chosen: for example, note that whenever the four satellites degenerate into a common plane [which happens 6 times per orbit, whenever $t_{n}=n\pi/12$, for odd $n$], they always form a perfect square containing the origin at its center.  

Since the first problem is static, the corresponding solution (\ref{q_hat}) has a static sort of symmetry: a group (the full tetrahedral group) of 24 spatial rotations and reflections that carry the configuration into itself.  Since the second problem is dynamical, the corresponding solution (\ref{p_hat}) has a dynamic sort of symmetry.  From a static standpoint, a ``still photograph" of (\ref{p_hat}) at some instant will, at most, be symmetric under 16 rotations and reflections (namely, the symmetry group $D_{4h}$ of the square prism, at the special times $t_{n}$ described above); but, from a dynamic standpoint, we can recall that the satellites' four angular momenta point along the four diagonals of the cube with vertices $\{\pm1,\pm1,\pm1\}$, and note that any rotation or reflection that leaves this cube invariant (there are 48 in total) also leaves the 4-satellite orbit (\ref{p_hat}) invariant, {\it when combined with an appropriate overall translation and/or reflection in time}.  In both solutions, (\ref{q_hat}) and (\ref{p_hat}), the four particles are equivalent to one another, in the sense that any particle may be mapped into any other by one or more of the symmetries.  While the symmetries of (\ref{q_hat}) are intuitively clear, the symmetries of (\ref{p_hat}) are considerably more subtle -- yet, as we have seen, (\ref{p_hat}) is actually more symmetrical than (\ref{q_hat})!  Just as (\ref{q_hat}) represents one of the simplest examples of a static lattice on the sphere, (\ref{p_hat}) represents one of the simplest examples of a choreographic lattice on the sphere.  Solution (\ref{q_hat}) was known in ancient times; but, as far as we can tell, (\ref{p_hat}) is new.

We believe that choreographic crystals share a beauty and naturalness with ordinary crystals, and we hope that they may be of similarly broad interest and importance.

\section{Symmetric Satellite Swarms}  

In the previous section, we introduced choreographic crystals via an example based on a symmetrical configuration of four satellite orbits.  In this section, we explain how to construct and classify {\it all} symmetrical satellite configurations.  (For earlier work on symmetric satellite configurations, see \cite{Mozhaev1, Mozhaev2}.)  In addition to being intrinsically (and technologically) interesting, this problem will set up our more general treatment of choreographic crystals in the subsequent section. 

Let us start by establishing some notation, terminology and conventions.  In this section, we can imagine for concreteness that each individual satellite moves on a Keplerian (elliptical) orbit with unit period.  The time translation operator $U_{\tau}$ maps each orbit $x=x_{i}(t)$ to the orbit $U_{\tau}x=(U_{\tau}x)_{i}(t)=x_{i}(t+\tau)$ (it shifts all of the satellites forward along their orbits by a common phase); the time reversal operator $T_{c}$ maps each orbit $x=x_{i}(t)$ to the orbit $T_{c}x=(T_{c}x)_{i}(t)=x_{i}(c-t)$ (it reverses all velocities and angular velocities, so the satellites move backward along their orbits); and an element $g=g_{ij}$ of the orthogonal group $O(3)$ maps each orbit $x=x_{i}(t)$ to the (rotated and/or reflected) orbit $gx=(gx)_{i}(t)=g_{ij}x_{j}(t)$.  We can also combine these operations: {\it e.g.}\ $(U_{\tau}gx)_{i}(t)=g_{ij}x_{j}(t+\tau)$ or $(T_{c}U_{\tau}x)_{i}(t)=(U_{-\tau}T_{c}x)_{i}(t)=x_{i}(c-t+\tau)$.  [To illustrate, consider two combinations -- (i) a $1/n$ time delay combined with a $2\pi/n$ rotation around the $\hat{z}$ axis, or (ii) zero time delay combined with a reflection through the $\{x,y\}$ plane: either combination leaves a circular orbit in the $\{x,y\}$ plane invariant, but acts non-trivially on an orbit which is tipped out of the $\{x,y\}$ plane.]

We refer to a set of satellite orbits as a ``swarm" ${\cal S}$.  A ``symmetry" (or ``symmetry operation" \cite{BurnsGlazer}) of the swarm ${\cal S}$ is a combined transformation $U_{\tau}g$ that leaves ${\cal S}$ invariant: ${\cal S}=\{U_{\tau}gx|x\in{\cal S}\}$; and a ``$\ast$symmetry" of the swarm ${\cal S}$ is a combined transformation $T_{c}U_{\tau}g$ that leaves ${\cal S}$ invariant: ${\cal S}=\{T_{c}U_{\tau}gx|x\in{\cal S}\}$.  In other words, a $\ast$symmetry involves time reversal, while a symmetry does not.  [For example, a swarm of circular orbits has a symmetry consisting of a time shift by half a period combined with spatial inversion through the origin (${\bf x}\to-{\bf x}$); and a single elliptical orbit has two $\ast$-symmetries: one which combines a time reversal with a rotation around the periapse direction, and another which combines a time reversal with a reflection in the plane spanned by the periapse and angular momentum directions.]  Let $G$ be a finite subgroup of $O(3)$: ${\cal S}$ is ``$G$-symmetric" if, for every $g\in G$, ${\cal S}$ has a symmetry of the form $U_{\tau}g$; and ${\cal S}$ is ``$G$-$\ast$symmetric" if, for every $g\in G$, ${\cal S}$ has a symmetry of the form $U_{\tau}g$ or a $\ast$symmetry of the form $T_{c}U_{\tau}g$.  Any $G$-$\ast$symmetric swarm is also an $H$-symmetric swarm, where $H$ is an index-2 subgroup of $G$ obtained by restricting to the symmetries of ${\cal S}$ that do not involve time reversal.  

The most basic type of $G$-symmetric swarm is a ``primitive $G$-symmetric swarm."  Every primitive $G$-symmetric swarm may be constructed in the following two steps.  First, choose a one-dimensional representation $\alpha$ of $G$; {\it i.e.}\ a function that maps each element $g\in G$ to a complex phase $\alpha(g)={\rm e}^{2\pi i\tau(g)}$, and satisfies $\alpha(g_1 g_2)=\alpha(g_1)\alpha(g_2)$.  Second, choose an integer $n$ and a fiducial satellite orbit $\bar{x}$ and construct the set of orbits ${\cal S}[G,\alpha,n,\bar{x}]=\{U_{[\tau(g)+m]/n}^{}g\bar{x}|g\in G,m\in\mathbb{Z}_{n}\}$.  If we take the union of two or more primitive $G$-symmetric swarms based on the same $G$, $\alpha$, and $n$ (but different fiducial orbits $\bar{x}_{1}$, $\bar{x}_{2}$, \ldots), we obtain another $G$-symmetric swarm, and any such swarm may be obtained this way.

The most basic type of $G$-$\ast$symmetric swarm is a ``primitive $G$-$\ast$symmetric swarm."  Every primitive $G$-$\ast$symmetric swarm may be constructed in the following three steps.  First, choose $H$, an index-2 subgroup of $G$, and let $g_{\ast}$ denote some (arbitrary but fixed) element of $G$ that is {\it not} in $H$: every element $g\in G$ may either be written as $g=h$ ($h\in H$) or as $g=g_{\ast}h$ ($h\in H$).  Second, choose a one-dimensional representation $\alpha$ of $H$, $\alpha(h)={\rm e}^{2\pi i\tau(h)}$, satisfying $\alpha(g_{\ast}h g_{\ast}^{-1})=\alpha(h)^{\ast}$ and
$\alpha(g_{\ast}^{2})=1$.  Third, choose an integer $n$ and a fiducial satellite orbit $\bar{x}$ and construct the set of orbits ${\cal S}[G,H,\alpha,n,\bar{x}]$ as the union of two sets: $\{U_{[\tau(h)+m]/n}h\bar{x}|h\in H,m\in\mathbb{Z}_{n}\}$ and $\{T_{c}U_{[\tau(h)+m]/n}g_{\ast}h\bar{x}|h\in H,m\in\mathbb{Z}_{n}\}$.  If we take the union of two or more primitive $G$-$\ast$invariant swarms based on the same $G$, $H$, $\alpha$, $n$, $g_{\ast}$ and $c$ (but different fiducial orbits $\bar{x}_{1}$, $\bar{x}_{2}$, \ldots), we obtain another $G$-$\ast$symmetric swarm, and any such swarm may be obtained this way.

For example, the 4-satellite orbit described in the previous section is a primative $G$-$\ast$symmetric swarm, where $G=O_{h}=\ast 432$ is the achiral octahedral group ({\it i.e.}\ the full symmetry group of the cube, including rotations and reflections); the index-2 subgroup is $H=T_{h}=3\ast 2$ (the pyritotetrahedral group); and the one-dimensional representation $\alpha$ of $H$ is $E_{u}$ (in Mulliken notation).

A primitive $G$-symmetric or $G$-$\ast$symmetric swarm is generated by acting on a fiducial satellite orbit $\bar{x}$ with $n|G|$ distinct operations (where $|G|$ is the order of $G$);  this process will generically produce a swarm with $n|G|$ satellites.  But if the fiducial orbit $\bar{x}$ and representation $\alpha$ are chosen carefully, then $\bar{x}$ will be invariant under some subgroup $K$ of the these operations, and we will instead generate a primitive $G$-invariant swarm in which the number of satellites is only $n|G|/|K|$.  Such orbits, in which an especially small number of satellites manage to represent an especially large number of symmetries, are of special interest and importance: the natural figure of merit here is $|K|/n$ or, equivalently, the total number of symmetries $|G|$ divided by the total number of satellites in the $G$-invariant swarm ${\cal S}$.  We call this number the ``choreography" $\chi$ of the swarm ${\cal S}$: a swarm ${\cal S}$ with large $\chi$ is like a delicately choreographed dance.  Since the finite subgroups $G\in O(3)$ ({\it i.e.}\ the ``3-dimensional point groups") have been completely classified \footnote{For a beautiful elementary and pedagogical introduction to the modern topological approach to this classification, due to Thurston and Conway, see Part 1 of ``The Symmetries of Things" \cite{Conway}.}, and the one-dimensional representations $\alpha$ of these groups are all known, it is straightforward to systematically sift through all possible swarms: we have done this, and found that that the swarm of highest choreography is precisely the 4-satellite configuration (\ref{p_hat}) introduced in the previous section, with choreography $\chi=12$ \footnote{This corresponds to the $G$-$\ast$symmetric lattice where $G=S_{4}\times \mathbb{Z}_{2}$ is the full octahedral group, $H=A_{4}\times \mathbb{Z}_{2}$ is the pyritohedral group, $n=1$, and $\alpha=v\times w$, where $v$ is one of the two non-trivial 1D representations of $A_{4}$, and $w$ is the non-trivial 1D representation of $\mathbb{Z}_{2}$.}.

\section{Generalizations}  

The symmetric satellite swarms in the previous section are a special case of a more general class of object with choreographic order.  It is natural to generalize the previous section in two different ways.

\begin{figure}[h]
  \begin{center}
    \includegraphics[width=3.4in]{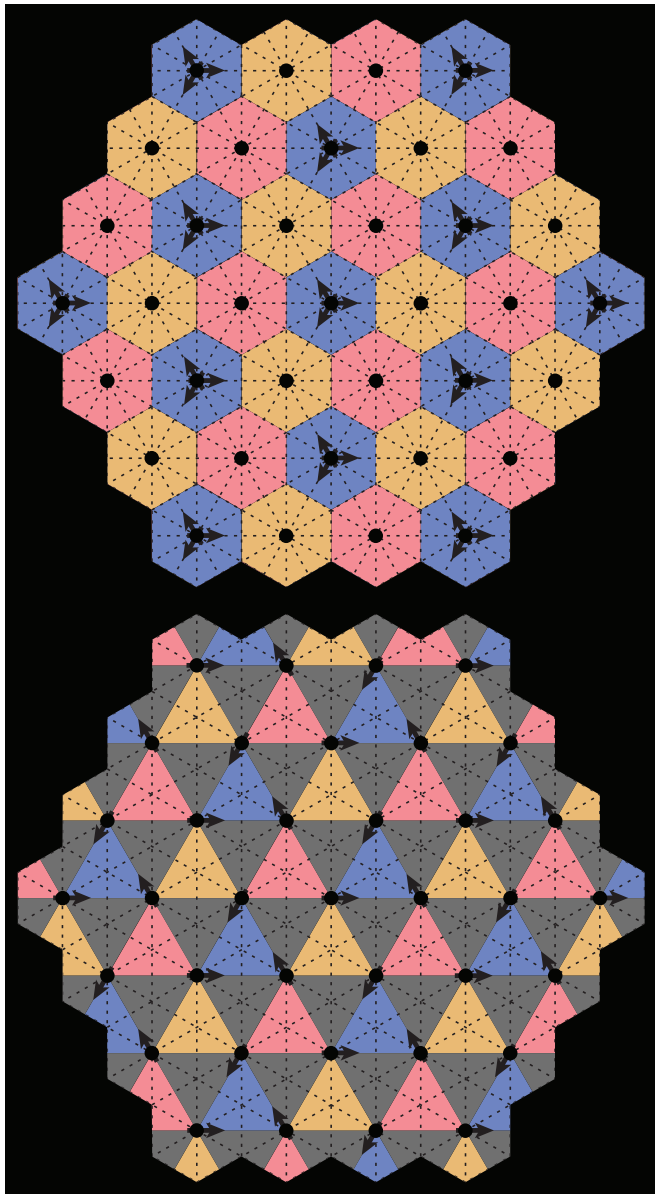}
  \end{center}
  \caption{The 2D planar choreographic crystal of highest choreography (top), 
    and another close contender (bottom).  Each arrow shows
    the initial position and velocity of a point that proceeds to move
    along a straight line ({\it i.e.}\ a geodesic in Euclidean space).
    Under each crystal, we have shown a colored tiling meant to help
    the reader see how the dance proceeds (from blue to yellow to
    pink, in repeating pattern).}
 \label{HexFigAandB}
\end{figure}

{\it Choreographic crystals on various geometries.}  Just as it is interesting to study static lattices, crystals and tilings on a wide variety of different spaces (the 2-sphere, the 3-sphere, 2D Euclidean space, 3D Euclidean space, \ldots), it is interesting to study choreographic order on a wide variety of spaces.  Choose an underlying space or space-time ({\it i.e.}\ an underlying Riemannian or pseudo-Riemannian geometry) ${\cal M}$ with isometry group ${\cal G}$; choose a discrete subgroup $G\in{\cal G}$; choose a one-dimensional unitary representation $\alpha$ of $G$: $\alpha(g)={\rm e}^{2\pi i\tau(g)}$  ($g\in G$); and choose a fiducial curve $\bar{x}=\bar{x}(t)$ in ${\cal M}$, where $t$ is a parameter along $\bar{x}$.  (Arguably the most natural case is when $\bar{x}$ is a geodesic of ${\cal M}$, and $t$ is an affine parameter, but we needn't restrict ourselves to this case.)  If $\bar{x}$ forms a closed loop, with $t$ varying over a finite range, we rescale it so that $0\leq t<1$; and if $\bar{x}$ is an infinite curve, with $t$ extending from $-\infty$ to $+\infty$, we distribute an infinite number of points along the curve (evenly spaced in $t$) and again rescale $t$ so that the spacing between successive points is unity.  Finally, we generate the ``primitive $G$-symmetric choreographic lattice" ${\cal C}[{\cal M},G,\alpha,n,\bar{x}]=\{U_{[\tau(g)+m]/n}g\bar{x}|g\in G,m\in\mathbb{Z}_{n}\}$.  The union of two of more such lattices based on the same ${\cal M}$, $G$, $\alpha$ and $n$ (but different fiducial orbits $\bar{x}_{1}$, $\bar{x}_{2}$, \ldots) is another $G$-symmetric choreographic lattice, and any such lattice may be obtained this way.  Alternatively, if $H$ is an index-2 subgroup of $G$, $g_{\ast}$ is an element in $G$ that is {\it not} in $H$, and $\alpha(h)={\rm e}^{2\pi i\tau(h)}$ is a one-dimensional unitary representation of $H$
satisfying $\alpha(g_{\ast}h g_{\ast}^{-1})=\alpha(h)^{\ast}$ and $\alpha(g_{\ast}^{2})=1$, then we may generate the ``primitive $G$-$\ast$symmetric choreographic lattice" ${\cal C}[{\cal M},G,H,\alpha,n,\bar{x}]$ as the union of the two sets $\{U_{[\tau(h)+m]/n}h\bar{x}|h\in H,m\in\mathbb{Z}_{n}\}$ and $\{T_{c}U_{[\tau(h)+m]/n}g_{\ast}h\bar{x}|h\in H,m\in\mathbb{Z}_{n}\}$.  The union of two or more primitive $G$-$\ast$invariant choreographic lattices based on the same ${\cal M}$, $G$, $H$, $\alpha$, $n$, $g_{\ast}$ and $c$ (but different fiducial orbits $\bar{x}_{1}$, $\bar{x}_{2}$, \ldots) is another $G$-$\ast$symmetric lattice, and any such lattice may be obtained this way.  

In this generalized context, we calculate the choreography $\chi$ as follows.  In a $G$-symmetric (or $G$-$\ast$symmetric) lattice, the isometry group $G$ ``folds" the underlying geometry ${\cal M}$ down to an irreducible patch or orbifold ${\cal O}$; or, in the other direction, the images of ${\cal O}$ under the action of $G$ give a natural tiling of ${\cal M}$.  The choreography $\chi$ is the number of orbifold tiles per point in ${\cal C}$ (or, equivalently, $|K|/n$, where $K$ is the stabilizer of $\bar{x}$).  This definition continues to be well-defined even when $|G|$ and the number of points in the ${\cal C}$ are infinite. 

An example should help bring the preceding formalism to life.  If we focus on the simple case where the background geometry ${\cal M}$ is the two-dimensional Euclidean plane, and the fiducial trajectory $\bar{x}$ is a geodesic in the plane (a straight line, with a particle moving along it at constant velocity), then the top panel in Fig.~(\ref{HexFigAandB}) shows what we believe to be the choreographic lattice of highest choreography ($\chi=12$), while the bottom panel shows another choreographic lattice with $\chi=6$.

{\it Generalized choreographic order.}  The choreographic lattices we have constructed thus far have come from simultaneously letting the isometry group $G\in {\cal G}$ of the background geometry ${\cal G}$ act on the orbits in two different ways -- ``directly" ($\bar{x}\to g\bar{x}$) and via a one-dimensional unitary representation ($\bar{x}\to U_{\tau(g)}\bar{x}$) -- and taking advantage of the interplay between these two actions.  There are other interesting possibilities in this direction which make use of other higher dimensional representations.  As a first example, imagine a swarm ${\cal S}$ in which the individual particles are not featureless satellites, but rather spin $s$ particles: then it would be natural to consider systems generated by letting the isometry group $G\in O(3)$ simultaneously act in three different ways: via the direct action (on the orbit's orientation, $\bar{x}\to g\bar{x}$), via a one dimensional unitary representation (that acts on the particle's orbital phase, $\bar{x}\to U_{\tau(g)}\bar{x}$), and via a $(2s+1)$ dimensional representation (that acts on the particle's spin).  From a quantum mechanical standpoint, one might also consider $N$-dimensional representations of $G$ that entangle collections of $N$ particles by mixing them (or their associated wave functions) at the same time as the isometry group acts on the underlying geometry directly.  It seems that many interesting forms of generalized choreographic order may be possible here.

\section{Diffraction Signatures}  An important open question is whether any actual many-body systems exhibit choreographic order, either in their ground state \footnote{in which case they would be examples of so-called ``time crystals" \cite{Shapere:2012nq, Wilczek:2012jt} (see also \cite{Li2012, BrunoComment1, BrunoComment2, Wilczek:2013uca, BrunoNoGo, Wilczek:2013})}, or when appropriately prepared and/or driven.  Whether or not such systems occur naturally, it should be possible to engineer them in the lab.  In either case, they should exhibit distinctive signatures in diffraction experiments, as we shall now explain.

{\it Modified Bragg Law.}  To get the idea, first recall that, in ordinary Bragg diffraction (from a static crystal), the diffraction peaks obey two rules \cite{AshcroftMermin}: (i) the difference $\Delta\vec{k}\equiv\vec{k}_{f}-\vec{k}_{i}$ between the initial and final wave vector is a point in the crystal's reciprocal lattice, while (ii) the difference $\Delta\omega\equiv\omega_{f}-\omega_{i}$ between the initial and final frequency vanishes.  In the case of choreographic order, the crystal may be divided into $N$ congruent sub-lattices, each moving with a different velocity.  (For example, the lattices in Fig.~\ref{HexFigAandB} may be decomposed into three sub-lattices with different velocities.)  We can calculate the diffraction separately for each sub-lattice, and then superpose the results.  The diffraction due to sub-lattice $\beta$ has the following properties: $\Delta\vec{k}$ is a point in the sub-lattice's reciprocal lattice, while $\Delta\omega$ is {\it non-vanishing}, and given by $\Delta\omega=\vec{v}_{\beta}\cdot\Delta\vec{k}$, where $\vec{v}_{\alpha}$ is the velocity of the sub-lattice \footnote{One can see this in two different ways: (i) by thinking about the doppler effect due to the motion of the lattice points in the lab frame; or (ii) by noting that each sub-lattice will satisfy the ordinary Bragg law in its reference frame, which will must be transformed to the lab frame.}.  Let us see the effect of this modified Bragg law in two standard experimental configurations \cite{AshcroftMermin}: von Laue diffraction and powder diffraction.

{\it von Laue diffraction and powder diffraction.}  In von Laue diffraction, a beam of particles of mass $m$ (with a range of energies, but a single fixed direction $\hat{k}_{i}$) is scattered off a single crystal of fixed orientation.  First focus on a particular diffraction peak due to sub-lattice $\beta$ (corresponding to a particular point $\vec{K}$ in its reciprocal lattice).  If sub-lattice $\beta$ were at rest, this peak would lie in the direction $\hat{k}_{f}$, with wavenumber $\bar{k}$ and frequency $\bar{\omega}=(m^{2}+\bar{k}^{2})^{1/2}$; but when we give the sub-lattice a small velocity $\vec{v}_{\beta}$, the direction and frequency of the peak changes (to first order in $\vec{v}_{\beta}$) as follows: the perturbed peak still lies in the unperturbed scattering plane spanned by $\hat{k}_{i}$ and $\hat{k}_{f}$, but the scattering angle $\theta$ shifts from its unperturbed value ${\rm cos}\,\theta_{0}=\hat{k}_{i}\cdot\hat{k}_{f}$ to the perturbed value ${\rm cos}\,\theta={\rm cos}\,\theta_{0}-(\vec{v}_{\beta}\cdot\vec{K})(\bar{\omega}/\bar{k}^{2})(1+{\rm cos}\,\theta_{0})$, while the frequency shifts by $\delta\omega_{f}=(\vec{v}_{\beta}\cdot\vec{K}){\rm cos}\,\theta_{0}/({\rm cos}\,\theta_{0}-1)$.  When we superpose the diffraction pattern from the $N$ different sub-lattices, we see that for the most part, the peaks from different sub-lattices do not overlap, but instead group into $N$-tuplets with small angular and frequency splittings described by the preceding formulae.  But for certain values of $\vec{K}$, it can happen that $\vec{v}_{\beta}\cdot\vec{K}$ and $\vec{v}_{\gamma}\cdot\vec{K}$ are the same, for two different sub-lattices $\beta$ and $\gamma$: in this case, these two peaks will interfere with one another, with the interference phase $(\vec{x}_{\beta}-\vec{x}_{\gamma})\cdot\vec{K}$, where $\vec{x}_{\beta}$ and $\vec{x}_{\gamma}$ are arbitrarily chosen points in sub-lattices $\beta$ and $\gamma$, at some arbitrary time $t$.  In powder diffraction, a beam with a single fixed energy $\bar{\omega}$ and direction $\hat{k}_{i}$ is scattered off a powder made up of crystals with all possible orientations.  In this case, when we give sub-lattice $\beta$ a small velocity $\vec{v}_{\beta}$, the scattering angle $\theta$ is shifted from its unperturbed value by $\theta_{0}$ to ${\rm cos}\,\theta={\rm cos}\,\theta_{0}+(\vec{v}_{\beta}\cdot\vec{K})(\bar{\omega}/\bar{k}^{2})[1-{\rm cos}\,\theta_{0}]$, while the unperturbed frequency $\bar{\omega}$ is shifted by $\delta\omega_{f}=\vec{v}_{\beta}\cdot\vec{K}$.  Otherwise, the story is the same as in the von Laue case: the peaks group into $N$-tuplets, with small splittings given by the preceding formulae; and interference when $\vec{v}_{\beta}\cdot\vec{K}=\vec{v}_{\gamma}\cdot\vec{K}$, with interference phase $(\vec{x}_{\beta}-\vec{x}_{\gamma})\cdot\vec{K}$.  

The possible crystals in 2D and 3D Euclidean space, and their corresponding diffraction patterns, will be explored further in subsequent work \cite{FollowUp}.   In the future, it will be interesting to reconsider the vibrational modes of an ordinary crystal, by thinking of them as a special type of choreographic crystal; or to explore the possible connections with previous work on generating higher harmonics of laser fields (which also relies crucially on the combined space-time symmetries of the system in question) \cite{Tong, Alon1, Alon2, Alon3, Alon4, Ceccherini, Alon5, Chu}; or to consider the possibility of choreographic {\it quasi}crystals.  It would, of course, be wonderful to engineer an example of a choreographic crystal in the lab, and even more wonderful to find a condensed matter system that has intrinsic choreographic order.  

We thank Dmitry Abanin, Paul Steinhardt and Xiao-Gang Wen for discussions, and the anonymous PRL referees for their valuable comments.  Research at the Perimeter Institute is supported by the Government of Canada through Industry Canada and by the Province of Ontario through the Ministry of Research \& Innovation.  LB also acknowledges support from an NSERC Discovery Grant.

\end{document}